\documentclass{article}
\usepackage{amsfonts,graphicx,bm,avant}
\makeatletter
\renewcommand\section{\@startsection {section}{1}{\z@}%
                                   {-3.5ex \@plus -1ex \@minus -.2ex}%
                                   {3ex \@plus.2ex}%
                                   {\normalfont\large\sf\bfseries\centering}}
\makeatother

\title{\sf Some questions of Monte-Carlo modeling
on nontrivial bundles}
\date{\sf 17 June 2007}
\author{\em Alexander Yu. Vlasov\thanks{
\tt Federal Radiology Center, IRH \hfil
197101, Mira Street 8, St.-Petersburg, Russia}}
\begin{document}
\sloppy
\newcommand{\ve}[1]{\boldsymbol{#1}}
\newcommand{\R}{{\mathbb R}}
\newcommand{\Sph}{{\mathbb S}}
\maketitle
\begin{abstract}
 In this work are considered some questions of Monte-Carlo modeling
on nontrivial bundles. As a basic example is used problem of generation
of straight lines in 3D space, related with modeling of interaction
of a solid body with a flux of particles and with some other tasks.
Space of lines used in given model is example of nontrivial fiber
bundle, that is equivalent with tangent sheaf of a sphere.
\end{abstract}

\section{Introduction}

Monte-Carlo method, or {\em method of statistical trials} is
a numerical method based on simulation by random variables and 
the construction of statistical estimators for the unknown quantities
\cite{slov}.
This method is used wider and wider due to permanent growth of capabilities
and accessibility of computers and sometime is considered as
a ``brute force method'' in comparison with more traditional
analytical and numerical methods.

Such a critical view is not always justified and in the presented
work is considered an example of practical task, when application of
Monte-Carlo method is very naturally related with differential
geometry and theory of fiber bundles.

One simplest application of Monte-Carlo method --- is computation
of multiple integrals. If it is required to estimate such an integral 
with respect to the Lebesgue measure in an $n$-dimensional Euclidean space, 
the generation of sequences of random points, distributed in such
a space does not produce some specific problems.

As a basic example with a compact support may be considered
the generation of uniformly distributed points in multidimensional
rectangular area, represented as the direct product of $n$ intervals
and containing given compact area. For generation of points
with more general distribution laws may be used quite standard
methods \cite{sprav}.

Such an approach does not produce especial difficulties, because
a {\em natural measure} exists on Euclidean space and it specifies
method of generation of uniformly distributed points, that may
be used as a base for creation of more general distributions,
produced via different transformations of the initial one.

It is more difficult to apply Monte-Carlo method to specific
tasks, related with distributions of lines, planes, or other
geometrical objects. A way to introduce a natural measure for 
such objects is not always obvious \cite{Ken}.

A classical example  --- is {\em Bertrand paradox} \cite{Ken},
associated with name of the French mathematician of XIX century (quite
detailed discussion may be found also in \cite[\S 19]{Gard}). It is
good illustration of considered class of problems. 

The Bertrand problem is stated as finding {\em probability for length
of random chord in circle to be bigger than side of equilateral
triangle, inscribed in given circle}. It is called ``paradox,''
because alternative methods of solutions of given task produce
different values for the probability. It is related with fact,
that it is necessary first to define meaning of term ``random''
chord \cite{Ken}.

\section{Straight lines in the space}

The statement of the problem, illustrated in the example with
Bertrand paradox, may be quite important for modeling of wide
range of physical problems using Monte-Carlo method. Let us
consider three-dimensional space $\R^3$ with a solid body and a
distribution of random lines, corresponding to trajectories of
particles, intersecting of given object.

It was already mentioned above, that it is necessary to describe
precisely, which particular distribution of lines denoted as
``random'' and in presented work as basic example is considered
{\em isotropic uniform distribution of lines}, defined by
suggestion, that any points of space and any directions are
equiprobable.

Despite of apparent simplicity of such definition, the example
with Bertrand paradox shows necessity of accurate consideration.
Say in Ref.~\cite{Ken} is suggested a general approach to description
of geometric objects: first, to choose appropriate parametrization
({\em co-ordinate system}) for description of geometric object
by {\em unique} way, and next, to define probability density
function ({\em measure}) for given space of parameters.

As an example of given approach in Ref.~\cite{Ken} is considered
description of a straight line in the space using equations:
\begin{equation}
\left\{
\begin{array}{rclcl}
 x &=& a\,z &+& p \\
 y &=& b\,z &+& q 
\end{array}
\right. .
\label{z0lin}
\end{equation}
Here $(a,b,p,q)$ --- are four parameters describing the straight line in
three-dimensional space $(x,y,z)$.

Invariant measure for considered parametrization $(a,b,p,q)$
Eq.~(\ref{z0lin}) may be defined as \cite{Ken,San}:
\begin{equation}
 (1+a^2+b^2)^{-2} da\,db\,dp\,dq.
\label{z0mes}
\end{equation}

It should be mentioned, that parametrization Eq.~(\ref{z0lin})
is not complete, because may not describe a set of lines
those parallel to plane $z=0$.

\begin{figure}[htb]
\begin{center}
\includegraphics[scale=0.5]{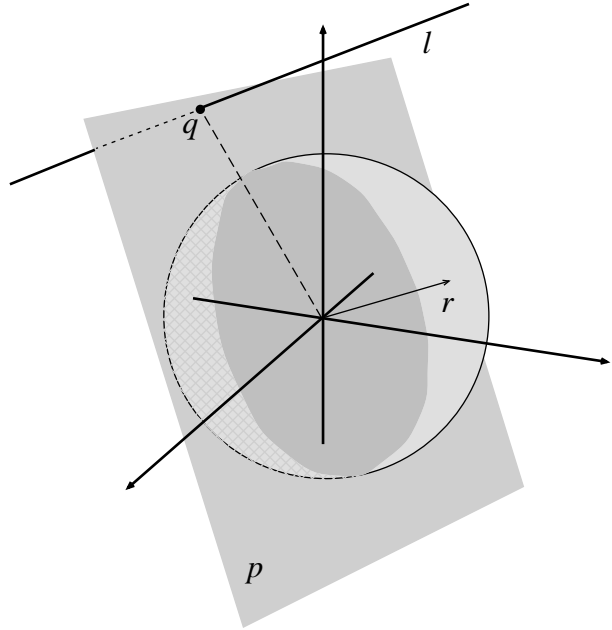} 
\end{center}
\caption{A line and a unit sphere of directions}\label{Fig:line}
\end{figure}

For complete definition of manifold of straight lines in
the space it may be used other method: to consider a plain $p$
containing origin of the co-ordinate system, point $q$ on the plane,
and to draw straight line $l$ perpendicular to the plane
through given point \cite{San,Amb} (see Fig.~\ref{Fig:line}).

This definition also specifies the space of lines as a 
four-dimensional manifold, because may be represented as a 
fiber bundle with two-dimensional base and two-dimensional 
fiber \cite{Amb}.

Really, any plane used in the definition may be presented by
a straight line drawn through the origin of co-ordinate system
and perpendicular to given plane. A space of such lines --- 
is the projective space\footnote{The projective space $\R P^n$
is space of rays $\ve{r} \sim \lambda\ve{r}$, 
$\ve{r} \in \R^{n+1}\setminus\{\ve{0}\}$, $\lambda \in \R$.} 
$\R P^2$, it is two-dimensional base of 
the fiber bundle. The two-dimensional fiber --- is the plane itself.

Maybe more visual example is {\em space of directed lines},
corresponding to tangent sheaf of the sphere, where base is
usual sphere and fibers --- are tangent planes to the sphere
\cite{Amb}. 

\section{Monte-Carlo method for space of lines}

In applications with Monte-Carlo method, last definition
produces following sequence of actions: to generate a unit
vector (radius $\ve{r}$, Fig.~\ref{Fig:line}); to draw the perpendicular 
plane $p$ through the origin of the coordinate system using the vector; 
to generate point $q$ on the plane; to draw (directed) line $l$ passing 
through given point and parallel to the vector.

Such a definition is convenient for modeling of isotropic uniform 
distribution of lines, because due to the symmetry reasons the both
distribution of directions of the radii (points on sphere) and 
distribution of points on plane should be uniform.

This example displays specific properties, because tangent sheaf
to the sphere (used for parametrization of space of straight
lines), {\em is not the trivial bundle}. So generation of
distributions using method described above may encounter
some difficulties. 

Really, straightforward approach could be described using
procedure with two steps: to generate point uniformly
distributed on unit sphere, to generate point uniformly
distributed on plane (more exactly, on compact subset  
with size sufficient for modeling of particle flux
for given volume).

Such a method meet a problem, because nontrivial bundle
may not be presented as direct product of base on fiber
and in general case probability distributions on fiber 
and base are not define distribution on total space.  

The example with tangent bundle to the sphere is quite
illustrative: a point on a tangent plane could be generated
using two coordinates, if global fibration of tangent 
co-ordinate frames on sphere is given. 
But it is impossible to do it in some regular ({\em continuous})
way, because on the sphere may not exist a field of unit 
tangent vectors.
It is so-called {\em the hedgehog theorem}\footnote{Also known
as the {\em hairy ball theorem}.} \cite{PostCel,Brou} 
(it is not possible to ``comb'' a hedgehog (ball), {\em i.e.}, to arrange all
spines smoothly).
 
\medskip

In particular case of modeling of isotropic uniform distribution of lines 
using Monte-Carlo method there are few ways to avoid the problem. Say, it
is possible to generate random tangent vector (or basis) for each plane.
It is possible, because due to symmetry of the model, orientation of
the basis does not matter. One method of construction of such random 
unit vector for arbitrary plane is: to generate random point in unit
ball, to find projection on given plane and to normalize it on
unit length.

From the one hand, this example shows specific advantages of 
Monte-Carlo method: it may be used even without proper parametrization
of space of integration. For usual analytical methods it could appear
necessity to use more difficult and indirect procedures, {\em e.g.},
construction of atlas with few maps and transition functions for 
description of a fiber bundle \cite{dif1} or deletion of part
of space, similar with parametrization Eq.~(\ref{z0lin}) above.

On the other hand, it would be convenient to have for arbitrary
distribution on space of lines some general and regular procedure
for Monte-Carlo method without suggestion about uniformity and
isotropy of distribution of straight lines.  

\medskip

Let us consider fiber bundle used in given model with
more details. It is convenient for simplicity again to
use tangent sheaf of the sphere, corresponding to space of
{\em directed} lines instead of initial fiber bundle over 
projective space.

It is known, that {\em any fiber bundle may be defined as
associated with some principal bundle} \cite{mod}, so it is
convenient first to consider examples with principal bundles
relevant to given problem.

By definition, {\em the fiber of principal bundle --- is Lie 
group}, acting freely (without fixed points) on the space
of the bundle. Two simple examples of principal bundles over
the sphere \cite{mod} are {\em frame bundle} (it corresponds to
a bundle of unit tangent vectors already mentioned before)
\begin{equation}
 \R P^3 \cong SO(3) \stackrel{p}{\longrightarrow} SO(3)/SO(2) \cong \Sph^2
\label{reps}
\end{equation}
with fiber is $SO(2) \cong \Sph^1$ and Hopf fibration \cite{mod,Ste}
\begin{equation}
 \Sph^3 \cong SU(2) \stackrel{p}{\longrightarrow} SU(2)/U(1) \cong \Sph^2
\label{Hopf}
\end{equation}
with the same fiber $U(1) \cong \Sph^1$, represented as subgroup of 
diagonal matrixes in $SU(2)$ and base $\Sph^2$, but with different 
total space, (hyper)sphere\footnote{Here $\Sph^n$ always denotes subspace
or unit vectors in $\R^{n+1}$, {\em i.e.}, $\Sph^2$ is usual sphere, 
$\Sph^1$ is circle and $\Sph^3$ is three-dimensional manifold, represented 
as unit hypersphere in $\R^4$.} $\Sph^3$. 

The both cases are examples of {\em principal bundles on homogeneous
spaces}, when space of bundle $G$ is Lie group, fiber $H$ is
subgroup and base --- is quotient space $M=G/H$. In such a case
action of element of group $h \in H$ on space $G$ is expressed
simply as {\em right} multiplication $G \ni g \mapsto gh$ \cite{dif1}.

One property of nontrivial principal bundles --- is absence of
continuous sections, {\em i.e.}, maps from base to total space
\cite{dif1,Ste}. Say, in the case with unit tangent vectors
Eq.~(\ref{reps}), such a map would produce tangent direction in
any point of sphere and so the {\em hedgehog theorem}
mentioned earlier is direct consequence of impossibility
of such (continuous) section.

It is the demonstration of a principle, that some problems of 
Monte-Carlo modeling, discussed it presented work
are really direct consequences of topological properties
of fiber bundles.

\smallskip

Let us consider a problem of Monte-Carlo modeling of distribution of
random tangent unit vectors on a sphere. For example, it is
necessary to generate random direction and point on surface
of ball with given probability distribution.

The fiber bundle is not trivial, and so for anisotropic distribution
it is not simple not only resolve, but even state the problem, if
to start from two independent distributions: points on sphere
and points on circle (direction). It is an analogue of situation
described above --- it is not possible to build continuous
global tangent coordinates on the sphere and so after generation
of random point on a sphere and random direction (point on circle)
it is not clear, how to ``attach'' given direction to given
point on sphere (``Where is a North on the South Pole?'').
 
However, such a problem may be simple resolved, if to start with
total space of the bundle. It is more correct way to state
the tasks for Monte-Carlo method for such a cases. Say,
for modeling of unit tangent vectors on a sphere, it is
necessary to generate random element $A \in SO(3)$ (rotation).

Really, such an element has one-to-one correspondence with
a frame $(\ve{x}',\ve{y}',\ve{z}')$, produced via rotation $A$
from fixed standard co-ordinate frame $(\ve{x},\ve{y},\ve{z})$.
Now, using the random (dashed) frame, it is possible to
simply resolve given task: let $\ve{z}'$ determines point on
sphere, then $\ve{x}'$ may be used as random direction from
this point.

Though this method assumes possibility to specify distributions 
and generate random elements on the space $\R P^3 \cong SO(3)$.
For calculations and applications of Monte-Carlo method it is
more convenient to start with space $\Sph^3 \cong SU(2)$. 
The projective space $\R P^3$ may be produced from hypersphere
$\Sph^3$ via identifying of all opposite points
$\R P^3 \cong \Sph^3/ \{1,-1\}$ and, similarly, there is
$2 \to 1$ covering, homomorphism of groups, then the same
rotation from group $SO(3)$ corresponds to two unitary
$2 \times 2$ matrixes: $a$ and $-a$ \cite{PostLie}
\begin{equation}
 SO(3) \cong SU(2) / \{1,-1\}.
\label{SU2SO3}
\end{equation}

For treatment of the case with $\R P^3 \cong SO(3)$ it is enough
to consider distributions of random points on hypersphere $\Sph^3$ with
densities satisfying property $\mu(\ve{r}) = \mu(-\ve{r})$, where $\ve{r}$ 
is unit 4-vector, describing hypersphere in four-dimensional space.

\medskip

Constructions are more difficult for {\em associated bundles}.
Such a bundle is produced by following method \cite{dif1}: if there is
principal bundle $P$ with base $M$ and fiber (structure group) $G$,
then for construction of associated bundle with the same base and
fiber $F$ (with left action of $G$), the direct product $P \times F$ 
with action $(u,f) \mapsto (ug,g^{-1}f)$ is defined first, and 
it is considered quotient space with respect to such action
$E = P \times_G F$. Now it is possible to represent $E$, as fiber
space of bundle with base $M$ and fiber $F$ \cite{mod,dif1}.

An example of the associated bundle --- is tangent sheaf to the sphere, 
used for description of space of (directed) lines. Formally, the construction
described above may be used directly for description of arbitrary
distributions on the tangent bundle of sphere.

Let us consider first the direct product $P \times F = SO(3) \times \R^2$.
It is set of pairs $(g,r)$, $g \in SO(3)$, $r \in \R^2$. It is necessary
now to introduce space $E$, {\em i.e.}, quotient space on equivalence
relation $(g,r) \sim (gh,h^{-1}r)$, there $h$ is element of group $SO(2)$,
that is embedded in $SO(3)$ as subgroup of rotations around a fixed
axis (say $\ve{z}$), on plane $SO(2)$ acts as rotations around origin.

The quotient space 
\begin{equation}
 E = SO(3) \times_{SO(2)} \R^2
\label{SSR}
\end{equation}
has quite difficult
structure and so for modeling of distributions on this space it is
possible to use a method with invariant measure, already used earlier.
Namely, any distribution on four-dimensional manifold $E$ may be 
identified with distribution on five-dimensional manifold 
$SO(3) \times \R^2$, with additional invariance property
\begin{equation}
\mu(g,r) \sim \mu(gh,h^{-1}r), \quad \forall\, h \in SO(2).
\label{mu5}
\end{equation}

For applications to Monte-Carlo method such an algorithm may be described 
as follows: to generate random element of five-dimensional set 
$(A,r) \in SO(3) \times \R^2$ with distribution satisfying Eq.~(\ref{mu5}). 
The first element, $A \in SO(3)$, produces random frame $(\ve{x}',\ve{y}',\ve{z}')$ 
and it was already discussed earlier in example with random tangent vector.
The second element, $r \in \R^2$, describes point on plane. The first
element also describes orientation of the plane via pair $(\ve{x}',\ve{y}')$,
and it resolves the problem with lack of continuous tangent co-ordinate field 
on the sphere.

\section{Conclusion}
 
In this paper were considered some questions of Monte-Carlo modeling
on nontrivial bundles. Though as the basic example was used distribution
of straight lines in the space, it is possible to mention some general
principles, which were used in presented work and may be applied in
many other cases.

For consideration of nontrivial bundles it is useful to come from standard
description, as {\em projection from total space of bundle on a base} \cite{mod},
instead of less formal consideration of fiber bundle as some construction
with base and fiber (aka ``skew product'').

The definition of probability distribution on total space of the bundle
always produces an adequate method of description, but separate constructions
for distributions on base and fiber may be used either for trivial bundles
(direct product of base and fiber), or for a case with invariance of
distribution with respect to action of a structure group on the fiber,
similarly with example of isotropic uniform distribution of lines.

Such a method is appropriate for spaces of bundles with simple 
geometrical structure, because it is necessary to have possibility of 
application of numerical methods for generation of distributions on 
these spaces. An example of such fiber bundles is the principal bundle
on homogeneous spaces with not very complicated Lie groups, {\em e.g.},
Hopf fibration discussed above Eq.~(\ref{Hopf}).

Yet another method used in present work for description of {\em associated
bundles} --- is consideration of invariant densities Eq.~(\ref{mu5}).
This method also may be used in many other cases, because 
in definition of associated bundles is used quotient space
$E = P \times_G F$.

\smallskip

{\em Notes:} The presentation of the problem here should not be
considered as multifold. On the one hand, often may
be used simplified approach with modeling of isotropic uniform 
distribution of lines {\em inside of a sphere} via surface source 
on the sphere with cosine angular distribution \cite{Kel}. 
It is yet another example of incomplete parametrization 
of space of lines. 
On the other hand, instead of using associate bundle Eq.~(\ref{SSR}) 
space of lines may be presented directly as quotient space
of group $ISO(3)$ of all isometries of $\R^3$ \cite{Helg}
or described using theory of complex surfaces and twistors \cite{Hitch}.

\end{document}